\documentclass[twocolumn,aps,superscriptaddress,nofootinbib,floatfix,linenumbers]{revtex4}
\usepackage{epsfig,bm,feynmf}
\usepackage{graphics}
\usepackage{amsmath}
\usepackage{mathrsfs}
\usepackage{appendix}
\usepackage{bm}
\usepackage[normalem]{ulem}  
\usepackage[dvips]{color} 

\renewcommand{\sout}{\bgroup \color{red} \ULdepth=-.5ex \ULset}

\begin{document}
\title{Study of collective anisotropies $v_2$ and $v_3$ and their fluctuations in \textit{$pA$} collisions at LHC within a relativistic transport approach}

\author{Yifeng Sun}
\email{sunyfphy@lns.infn.it}
\affiliation{Laboratori Nazionali del Sud, INFN-LNS, Via S. Sofia 62, I-95123 Catania, Italy}

\author{Salvatore Plumari}
\email{salvatore.plumari@ct.infn.it}
\affiliation{Department of Physics and Astronomy, University of Catania, Via S. Sofia 64, 1-95125 Catania, Italy}
\affiliation{Laboratori Nazionali del Sud, INFN-LNS, Via S. Sofia 62, I-95123 Catania, Italy}

\author{Vincenzo Greco}
\email{greco@lns.infn.it}
\affiliation{Laboratori Nazionali del Sud, INFN-LNS, Via S. Sofia 62, I-95123 Catania, Italy}
\affiliation{Department of Physics and Astronomy, University of Catania, Via S. Sofia 64, 1-95125 Catania, Italy}

\date{\today}

\begin{abstract}
We have developed a relativistic transport approach at fixed $\eta/s(T)$ that incorporates initial space fluctuations generated by wounded quark model to study the hadron observables in 5.02 TeV p+Pb collisions. We find that our approach is able to correctly predict quite well several existing experimental measurements assuming a matter with $\eta/s=1/4\pi$, a result similar to previous studies within a viscous hydrodynamics approach.  Besides, we further discuss the sensitivity of the results on both $\eta/s(T)$ and the smearing width. Our transport approach has the possibility to include in initial conditions the power law tail associated to minijet, and this improvement extends the agreement with the experimental data to higher $p_T$ ranges. We also perform a comparison to Pb+Pb collisions pointing out that even if the collective flows have a similar magnitude the features of the matter created are different. By studying the correlation between collective flows and initial geometry, we find that the correlation decreases faster in small systems with the increase of $n$ and centrality. In particular we show that the variance of $\sigma_{v_n}/\langle v_n\rangle$ has a quite different evolution with centrality for p+Pb, so their measurement could provide some further hint about the correctness of current modelling.

t

\end{abstract}

\maketitle

\section{Introduction}
High energy nucleus-nucleus collision experiments at the BNL Relativistic Heavy Ion Collider (RHIC)
~\cite{Adams:2005dq,Adcox:2004mh,Arsene:2004fa,Back:2004je} and the CERN Large Hadron Collider (LHC)~\cite{Aamodt:2008zz} have provided the convincing evidence for the formation of a quark-gluon plasma (QGP) in the early stages of a heavy ion collision. This QGP has been found to be strongly correlated and to exhibit strong collective behavior. The theoretical calculations within viscous hydrodynamics~\cite{Romatschke:2007mq,Song:2008si,Schenke:2010rr,Gale:2013da,Song:2010aq,Niemi:2015qia} and transport approach~\cite{Ferini:2008he,Xu:2008av,Ruggieri:2013ova, Plumari:2015cfa,Konchakovski:2012yg,Karpenko:2012yf} have shown that such behavior is consistent with a matter with low shear viscosity to specific entropy ratio $\eta/s \simeq \, 0.1-0.2$ close to the conjectured lower bound $\eta/s=1/4\pi$ for a strongly interacting system~\cite{Kovtun:2004de}.

In recent years, experimental measurements from collisions of protons and deuterons with heavy nuclei reveal that high multiplicity events also show strikingly similar collective behavior as those seen in heavy ion collisions~\cite{CMS:2012qk,Abelev:2012ola,Aad:2012gla,Adare:2013piz}. Several theoretical studies based on hydrodynamics~\cite{Bozek:2011if,Bozek:2013uha,Kozlov:2014fqa,Shen:2016zpp,Weller:2017tsr} and transport approach~\cite{Bzdak:2014dia, Greif:2017bnr} show that this behavior can also be attributed to final state interactions (hydro-like expansion)  similarly to nucleus-nucleus collisions. 

We discuss here the results obtained within a relativistic transport approach where the collision integral is tuned to 
describe the evolution for a fluid at finite $\eta/s$. Such an approach has been shown to reproduce the viscous hydrodynamics
system expansion in $AA$ collisions \cite{Ruggieri:2013ova,Plumari:2015cfa} and even the ideal hydrodynamics behavior  in the infinite cross section limit \cite{Plumari:2015sia}. The advantage of the transport approach is that one can naturally include initial condition with a power law tail at $p_T>$ 2 GeV$/c$ representing the minijet distribution and follow the dynamics also at increasing 
Knudsen number (or large $\eta/s$). In this paper we in particular show a first study, to our knowledge, of the 
the correlations between the final collective flows and initial geometry, which can shed light on the understanding if the experimental measurements are due to final state interactions or some initial condition effects~\cite{McLerran:2016snu,Mace:2018vwq,Greif:2017bnr}. For this end, we thus develop an event-by-event transport approach that incorporates initial fluctuations to study the collective behaviors in p+Pb collisions, and the correlations between initial geometry and final collective flows. We have also found
in particular that the variance of $v_2$ and $v_3$ as a function of centrality should show a pattern
quite different with respect to the one in Pb+Pb collisions.

This paper is organized as follows. In the next section, we will detail the model setup of our kinetic approach, and how we  generate initial conditions event-by-event. In the section following, we will compare our results on hadron observables to experimental results after constraining the parameters related to the initial conditions. Sec. IV will discuss the correlation coefficient that characterizes the correlations between collective flows and initial geometry and the distribution of them. Finally, conclusions and discussions will appear in Sec. V. 

\section{Model setup}

To model the evolution of the small collision systems, we are employing the same relativistic transport code developed in these years~\cite{Ferini:2008he,Plumari:2011re,Plumari:2012ep,Ruggieri:2013bda,Ruggieri:2013ova,Plumari:2015cfa}, which has been used to study the dynamics of heavy-ion collisions at both RHIC and LHC energies, by solving the Relativistic Boltzmann Transport (RBT) equation. Numerically we solve the RBT equation using the test particle method, and the collision integral is solved by Monte Carlo method based on stochastic interpretation of transition amplitude~\cite{Xu:2004mz,Ferini:2008he,Plumari:2012ep}.
\begin{equation}
p_\mu \, \partial^\mu f(x,p)+ m^* \partial^\mu m^* \partial_{p^\mu}f(x,p)= {\cal C} [f]
\end{equation}
The key aspect of our approach is to gauge the collision integral ${\cal C} [f(x,p)]$ locally to a specific value of 
the viscosity to  entropy density ratio $\eta/s$ ~\cite{Plumari:2012ep}. Furthermore we can switch smoothly it to an increasing
$\eta/s$ one, reaching the estimated value in hadronic phase~\cite{Chen:2007xe,Demir:2008tr}, when the system reaches the cross over region. 
This is realized by determining the total isotropic cross section according to the Chapmann-Enskog approximation:
\begin{equation}
\eta = f(z) \frac{T}{\sigma_{tot}}
\label{eq:eta_M}
\end{equation}
where $z=m/T$ while the function $f(z)$ is defined by the following expression
\begin{eqnarray}
f(z) = \frac{15}{16}\frac{\left[ z^2 K_3(z)\right]^{2}}
 {(15z^2+2)K_2(2z)+(3z^3+49z)K_3(2z)}
\end{eqnarray}
where $K_n$-s are the modified Bessel functions. 
The entropy density for a massive system is given by $s=\rho \big( 4 + z K_1(z)/K_2(z) \big)$.
In the ultra-relativistic limit $z\to 0$ and the function $f(z) \to 1.2$ and we recover the massless limit for the $\eta$. 
As shown in \cite{Plumari:2012ep} 
The expression for $\eta$ in Eq.(\ref{eq:eta_M}) is in quite good agreement
at level of $3-5\%$ with the Green-Kubo formula \cite{Plumari:2012ep}. The final hadron is recovered at the end of the evolution using parton-hadron duality ansatz.

For initial conditions of partons, we use a modified Monte Carlo Glauber model assuming three constituent quarks localized within each nucleon inspired by wounded quark model~\cite{Bialas:1978ze,Anisovich:1977av,Eremin:2003qn,Bozek:2016kpf}, which can naturally obtain the linearity between the multiplicity of charged hadrons and the number of wounded quarks. 

For that reason, we firstly randomly place the constituent quarks in the nucleons, where the positions of nucleons in Pb nuclei are distributed according to the standard Woods-Saxon distribution with parameters $R_0=6.5$ fm and $a=0.54$ fm, according to the distribution $dN/dr=\frac{r^2}{r_0^3}e^{-r/r_0}$ with $r_0=0.3$ fm~\cite{Bozek:2016kpf}, and then shift the center of mass of the three quarks to the position of the nucleon. After that, we generate the wounded quark profile using Monte Carlo Glauber model, where we decide whether each quark pair from target and projectile can collide or not with a probability $p=e^{-\pi r^2/\sigma_{qq}}$ with $\sigma_{qq}=13.6$ mb in 5.02 TeV p+Pb collision~\cite{Bozek:2016kpf}.

\begin{figure}
\centering
\includegraphics[width=1\linewidth]{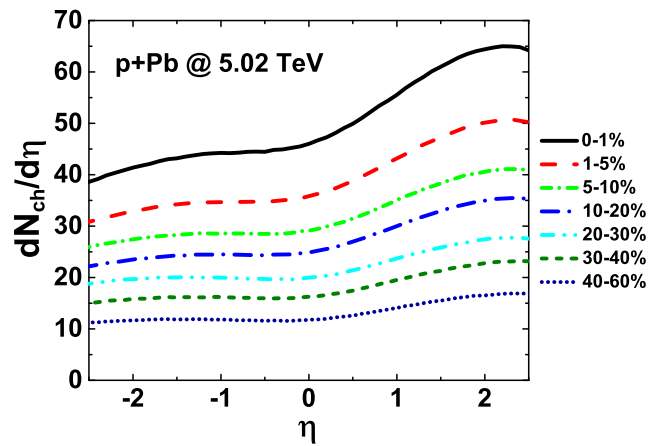}
\caption{(Color online) The pseudorapidity dependence of charged hadron multiplicity $dN/d\eta$ for various centrality bins in 5.02 TeV p+Pb collisions.}
\label{fig:dndeta}
\end{figure}

For the distribution of the spatial rapidity, we take the profile from Ref.~\cite{Bozek:2010bi}
\begin{equation}
\rho_{L\pm}(\eta)=(1\pm\frac{\eta}{\eta_m})\rm{exp}(-\frac{(|\eta|-\eta_0)^2}{2\sigma_{\eta}^2}\theta(|\eta|-\eta_0)),
\end{equation}
where the parameters are chosen as $\eta_m=5.7$, $\eta_0=2.5$ and $\sigma_{\eta}=2.5$ to have the same shape of $dN_{\rm{ch}}/d\eta$ measured by ATLAS Collaboration~\cite{Milov:2014ppp} as shown in Fig.~\ref{fig:dndeta}. This profile can account for more particles produced in the direction of nucleus, and greater asymmetry in events with larger multiplicity. 

After using the above procedure, the total initial parton density is then given as 
\begin{equation}
\frac{dN}{d^2\mathbf{x_{\perp}}d\eta}=\sum_{i=1}^{N_{\rm{part}}}n_i \rho_{\perp}(\mathbf{x_{\perp}}-\mathbf{x_i})\rho_{L\pm}(\eta),
\label{density}
\end{equation}
where $\mathbf{x_i}$ is the transverse position of each participant quark, $n_i$ is the number of partons generated by each participant, and $\rho_{\perp}(\mathbf{x_{\perp}})=\frac{1}{2\pi\sigma^2}e^{-\frac{\mathbf{x_{\perp}}^2}{2\sigma^2}}$. For the Gaussian distribution of the parton transverse density, we will change the parameter $\sigma$ in the range of 0.4-0.6 fm in order to 
study its effect on the integrated $v_2$ at each centrality. 

\begin{figure}[h]
\centering
\includegraphics[width=1.0\linewidth]{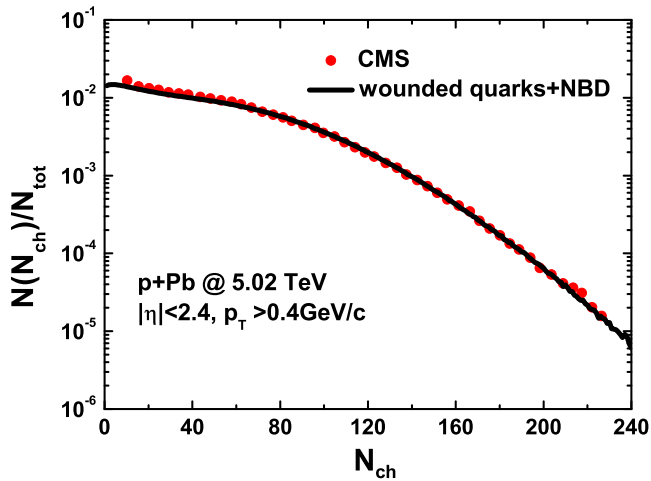}
\caption{(Color online) The multiplicity distribution of charged hadrons at $|\eta|<2.4$ and $p_T>0.4$ GeV$/c$ in 5.02 TeV p+Pb collisions from the MC Glauber model of wounded quarks supplemented with negative binomial distribution on each participant (wounded quarks+NBD), compared to CMS data~\cite{Chatrchyan:2013nka}.}
\label{fig:PNch}
\end{figure}

As the number of particles produced in p+p collisions fluctuates according to a negative binomial distribution (NBD), we thus take $n_i$ in Eq. (\ref{density}) to be $n_0N$, where $N$ is sampled according to NBD $P(N)=\frac{\Gamma(N+\kappa){\overline{n}}^N\kappa^{\kappa}}{\Gamma(\kappa)N!(\overline{n}+\kappa)^{N+\kappa}}$, and $n_0$ is a constant such that the final charged particle multiplicity is same as that measured in experiments. We find that $\kappa=0.54$, $\overline{n}=3.9$ and $n_0\approx2.352$ can almost reproduce the distribution of charged particle measured by CMS Collaboration~\cite{Chatrchyan:2013nka}, as shown in Fig.~\ref{fig:PNch}. We also mention that our $N_{\rm{ch}}$ is assumed to be equal to $N_{\rm{trk}}^{\rm{offline}}$ after the efficiency corrections usually adopted in hydro calculation~\cite{Kozlov:2014fqa,Shen:2016zpp,Bozek:2016kpf}. This however may introduce some uncertainty in selecting the beam centrality when comparing to data.
We notice that the values chosen are the same as the ones employed in an early approach to $pA$ collisions~\cite{Bozek:2016kpf}.

To get the momentum distribution of initial partons, we employ a blast wave model without initial transverse flow for it:
\begin{eqnarray}
&&\frac{dN}{d^2\mathbf{x_{\perp}}d\eta}=\frac{g\tau_0}{(2\pi)^2}p_Tm_T\nonumber
\\
&& \times\rm{cosh(\eta-y)}e^{-m_T\rm{cosh}(\eta-y)/T(\mathbf{x_{\perp}},\eta)}dp_Tdy,
\label{BW}
\end{eqnarray}
where $g=2\times8+3\times2\times6=52$ is the degree of freedom of partons (three flavor quarks and gluons), $\tau_0=0.4$ fm$/c$ is the thermalization time (again taken as in standard hydro approach for $pA$), and $m_T=\sqrt{m^2+p_T^2}$ is the transverse invariant mass with $m=0.3$ GeV. 
This value is chosen because it entails a correct asymmetry in the pseudorapidity charged particle distribution, as shown in 
Fig. \ref{fig:dndeta}. Also in thermal equilibrium it generates an equation of state close to the one calculated in lattice QCD \cite{Plumari:2019gwq}.
We use the above relation to calculate the temperature locally from initial parton density, and then sample the momenta of partons according to Eq. (\ref{BW}).

Finally, in order to compare our results to experimental data, we need to shift the rapidity of all charged hadrons from $y^*$ in center of mass frame to $y=y^*-0.465$ in lab frame, if we define the movement of Pb nuclei as the positive $z$ direction, as the beam energies were 4 TeV for protons and 1.58 TeV per nucleon for lead nuclei in 5.02 TeV p+Pb collisions.

\section{comparison to experiment}

Before comparing our results directly to measured data, we firstly need to check the convergence of our simulation. To do this, we use the modified Glauber model to generate the same initial conditions of partons, and then evolve the systems until kinetic freeze out.
This is the first time we employ our RBT approach to study $p$ collisions.
In fact the dimension and the densities explored by a $pA$ system are  significantly smaller than the ones in $AA$,
therefore we performed a convergency test to fix the grid size and the $N_{\rm test}$ appropriate to compute observables, where $N_{\rm test}$ is the number of test particle for each real particle,
and in particular $v_2$ and $v_3$, in $pA$ collisions.
We have found that  $N_{\rm{test}}=4000$ and $\Delta x=\Delta y=0.15$ fm in the simulations guarantee a convergency 
for $v_{2,3}(p_T)$ up to $p_T \sim $5 GeV$/c$.

\begin{figure}[h]
\centering
\includegraphics[width=1\linewidth]{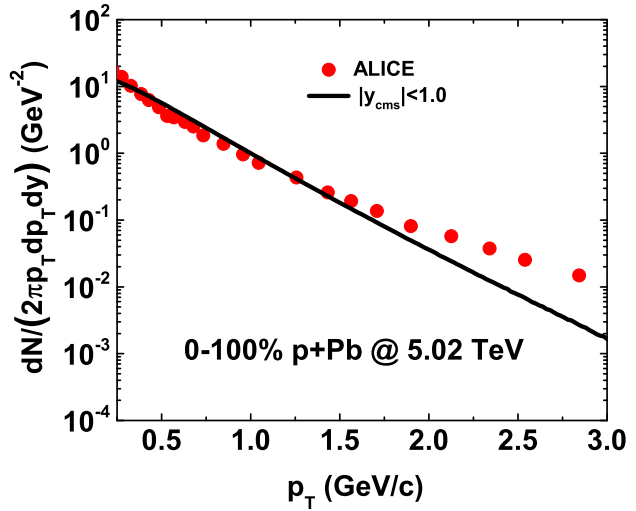}
\caption{(Color online) 
Charged hadron spectra at midrapidity in center of mass frame in minimum-bias 5.02 TeV p+Pb collisions, compared to data from the ALICE Collaboration~\cite{Abelev:2014dsa}.}
\label{fig:spectra}
\end{figure}

In Fig.~\ref{fig:spectra}, we compare our spectra to experimental measurements in minimum-bias 5.02 TeV p+Pb collisions from ALICE Collaboration~\cite{Abelev:2014dsa}, where the spectra is calculated in the center of mass frame of p+Pb collisions in both our studies and experiments. It is seen that our calculations agree with experimental ones in the range $0.3<p_T<1.5$ GeV$/c$, while they underestimate the charged particle multiplicity at both low and high $p_T$ due to two different reasons. The underestimation of charged particles at low $p_T$ can be attributed to the missing of resonance decays in our approach, which is absent in this study using simple parton-hadron duality ansatz. At higher $p_T$  we do not get a good description of the spectrum, however this feature is 
similar to the one in hydro approaches \cite{Bozek:2011if,Shen:2016zpp}. On the other hand the transport approach can be naturally extended including an initial
non-equilibrium distribution with the power law tail at increasing $p_T$ associated to the production of minijets.
We will see in the next paragraph that indeed the inclusion of minijets will allow to extend the validity region of the present approach
with respect to hydrodynamics, see Fig.\ref{fig:spectrajet}.

\begin{figure}[h]
\centering
\includegraphics[width=1\linewidth]{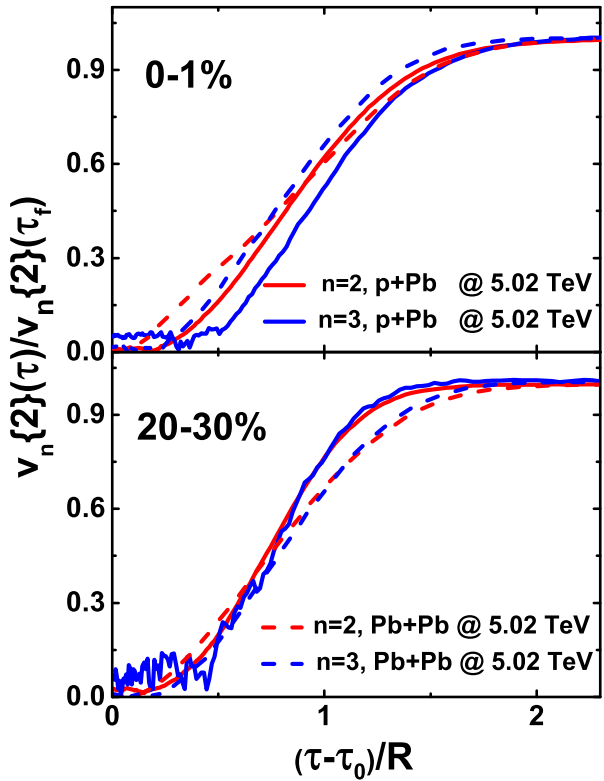}
\caption{(Color online) 
Time evolution of $v_2$ and $v_3$ as a function of time normalized to the mean square root size of the systems p+Pb (solid lines)
and Pb+Pb (dashed lines)
at two different centralities: 0-1\% (upper panel) and 20-30\% (lower panel).}
\label{fig:tvn-r}
\end{figure}

The significantly large anisotropies $v_{2,3}$ observed experimentally in p+Pb collisions had initially be seen as surprising,
because in $AA$ collision they were associated to formation time of about 4-5 fm/c while the high density state in p+Pb
collision is expected to be quite shorter. However in \cite{Bhalerao:2005mm} it was already discussed 
that the $v_2$ in an ideal hydrodynamical
expansion (zero viscosity) with a conformal equation of state can be expected to scale with the size of the formed system at different centralities.
We therefore investigated how the $v_2$ and $v_3$ evolve with time scaling the last with the mean square radius 
$\rm R=\sqrt{\langle r^2\rangle} $, where $\langle r^2\rangle$ is the mean square radius of the system weighted with the local density of the system.
Indeed the value of $\rm R$ goes from values of about 1 fm for p+Pb (almost independent on centrality) up
 to value of about 3.4 fm for Pb+Pb at  20-30\% and 4.6 fm in Pb+Pb at 0-1\% centrality.
In Fig. \ref{fig:tvn-r} we show by red lines the time evolution of the building up of $v_2$ 
(normalized to its asymptotic maximum value) in the upper panel in central collisions and in the lower panel in mid peripheral collisions for p+Pb (solid line) and
Pb+Pb (dashed line) and for $v_3$ (blue line).
We see that there is an approximate scaling with the time normalized to the size of the system R, and within
a time of about 1.5 R for $v_2$ and $v_3$ the anisotropies are already completely developed.

\begin{figure}[h]
\centering
\includegraphics[width=1\linewidth]{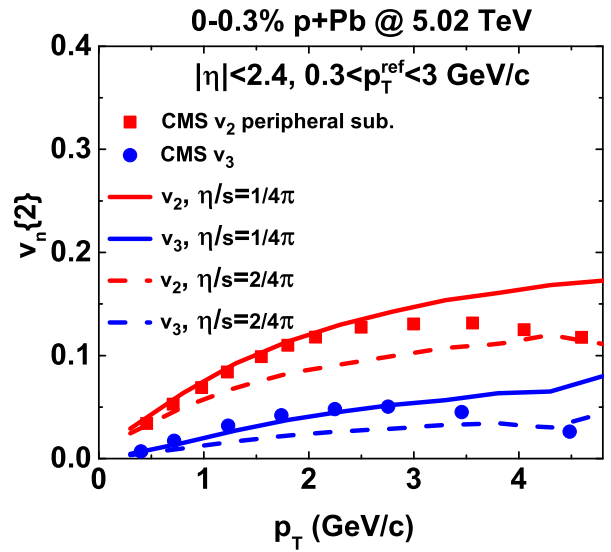}
\caption{(Color online) 
The transverse momentum $p_T$ dependence of $v_2$ and $v_3$ of charged hadrons at $|\eta|<2.4$ in 0-0.3\% 5.02 TeV p+Pb collisions for different choices of viscosity to specific entropy ratio $\eta/s$, compared to data from the CMS Collaboration~\cite{Chatrchyan:2013nka}.}
\label{fig:vn}
\end{figure}

Using the width $\sigma=0.55$ fm for gaussian fluctuations of the parton transverse density and $\eta/s=1/4\pi$, we find also a good agreement with experimental data measured by CMS Collaboration~\cite{Chatrchyan:2013nka} for elliptic and triangular flows, as shown in Fig.~\ref{fig:vn}. It is seen by the solid red and blue lines that $v_2$ and $v_3$ agrees with experimental data up to $p_T \simeq 2.5$ GeV$/c$ in our calculations which is quite similar to the results obtained within viscous hydrodynamics with very similar 
initial conditions \cite{Bozek:2011if,Bozek:2013uha,Shen:2016zpp}. We notice that we will compare our results with $v_2\{2\}$ after jet contribution subtracted. We recall that experimentally $v_2\{2\}$ is significantly larger than $v_2\{4\}$. Such an aspect should be considered more carefully in future studies especially in view of assessing to what extent the expansion of the matter created in $pA$ is of pure hydrodynamical nature as well as for a more precise determination of $\eta/s$. The disagreement between our calculations and experimental measurement at higher $p_T$ can be attributed to the missing of minijet production in higher transverse momentum, where non-equilibrium effect becomes more important. We also discuss this in next section. 

We also study the effect of $\eta/s$ on $v_2$ and $v_3$, which is shown by the dashed red and blue lines in Fig.~\ref{fig:vn}. It is seen that increasing $\eta/s$ to $2/4\pi$ can lower $v_2$ and $v_3$ for all $p_T$, though the effect is stronger in higher $p_T$, which also agrees with hydrodynamic approach.

\begin{figure}[h]
\centering
\includegraphics[width=1\linewidth]{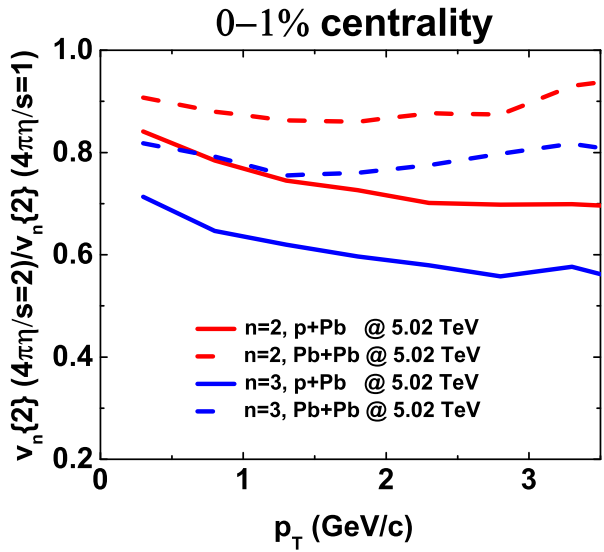}
\caption{(Color online) 
The ratio between $v_2$ (and $v_3$) with $\eta/s=1/4\pi$ and with $\eta/s=2/4\pi$ as a function of $p_T$ in 5.02 TeV p+Pb collisions and 5.02 TeV Pb+Pb collisions with same parameters and at 0-1\% centrality.}
\label{fig:vnratio}
\end{figure}

It is interesting to study the difference between the effect of $\eta/s$ on collective flows for small colliding systems and large ones.
To this end, we show in Fig.~\ref{fig:vnratio} the ratio $v_n\{2\} (4\pi\eta/s=2)/v_n\{2\} (4\pi\eta/s=1)$ as a function of transverse momentum in 5.02 TeV p+Pb and 5.02 TeV Pb+Pb collisions at 0-1\% centrality with same parameters. We find that in general increasing $\eta/s$ leads to the relative smaller effect on the decrease of $v_2$ and relative larger effect on the decrease of $v_3$ for all $p_T$. However, it is found that this decrease is stronger at larger $p_T$ in small colliding systems, while it is almost uniform for all $p_T$ in large colliding systems. We find that the effect of $\eta/s$ is stronger for p+Pb collisions, which can be seen by the solid red and blue lines in Fig.~\ref{fig:vnratio}.
In fact the effect in Pb+Pb is about a $10\%$ for $v_2$ (dashed red line) and $20\%$ for $v_3$ (dahsed blue line)
while for p+Pb is about twice as large already at $p_T \simeq$ 1 GeV$/c$ and further increase with $p_T$.
This suggest that ultra-central p+Pb collisions may supply more sensitivity for the determination of the $\eta/s$
w.r.t. to Pb+Pb collisions. However this larger sensitivity has been seen in our calculation only for the most central 
collisions, while at large p+Pb centrality above $20-30 \%$ (small multiplicity $N_{\rm{ch}}< 80$) the impact
of $\eta/s$ becomes similar to Pb+Pb collisions if not smaller.

\begin{figure}[h]
\centering
\includegraphics[width=1\linewidth]{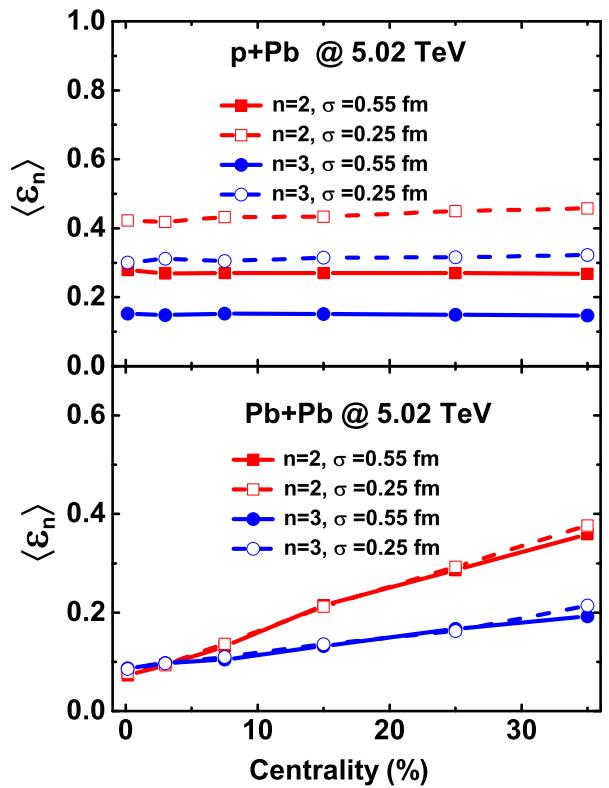}
\caption{(Color online) 
$\langle \epsilon_2\rangle$ and $\langle \epsilon_3\rangle$ as a function of centrality in p+Pb and Pb+Pb collisions for two cases: $\sigma=0.55$ fm and $\sigma=0.25$ fm.}
\label{fig:sigmae}
\end{figure}

To this end we have to add that the initial eccentricities in p+Pb collisions are strongly dependent on the width $\sigma$
of the fluctuation of transverse density, as it is shown in the upper panel of Fig. \ref{fig:sigmae}.
It is seen that there are dramatic enhancements of $\langle\epsilon_2\rangle$ and $\langle\epsilon_3\rangle$ with smaller $\sigma$.
Because $v_n$ and $\epsilon_n$ are strong correlated, we notice that in principle an increasing $\eta/s$ can be compensated
by a smaller $\sigma$ width of the spacial fluctuations that induce a larger $\epsilon_n$. However it is interesting to mention that a flat behavior of $\langle\epsilon_2\rangle \approx0.28$ with centrality is in nice agreement with the behavior extracted from a recent analysis by kinetic theory of pA collisions presented in Ref.~\cite{Kurkela:2019kip}. 

\begin{figure}[h]
\centering
\includegraphics[width=1\linewidth]{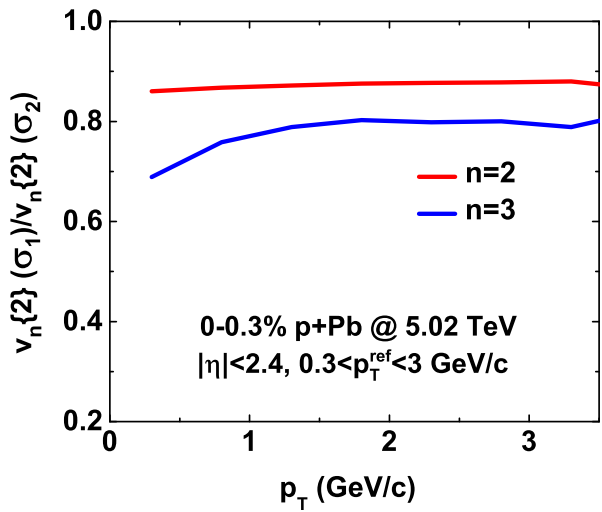}
\caption{(Color online) 
The ratio between $v_2$ (and $v_3$) with $\sigma=0.55$ fm  and with $\sigma=0.45$ fm as a function of $p_T$ in 5.02 TeV p+Pb collisions with same parameters and at 0-1$\%$ centrality.}
\label{fig:sigma12-ratio}
\end{figure}

In the lower panel of Fig.~\ref{fig:sigmae}, we also show $\epsilon_2$ and $\epsilon_3$ as a function of centrality in Pb+Pb collisions with $\sigma=0.55$ and $0.25$ fm; unlikely to p+Pb collisions, there is almost no dependence of eccentricity on $\sigma$. Therefore while in $AA$ collisions one has a very limited dependence on the width of the initial state fluctuations this
is large in $pA$ collisions. This limits the possibility to costrain the $\eta/s$ of the matter created if there are no independent
observables that allow to constraint also the initial state fluctuations. In Fig. \ref{fig:sigma12-ratio} we show the ratio of the anisotropies  as a function of transverse momentum
$v_n\{2\} (\sigma_1)/v_n\{2\} (\sigma_2)$ when changing the widths from a $\sigma_1=0.55 $ fm
to $\sigma_2=0.45 $ fm in ultra-central collisions(0-0.3\%), assuming an $\eta/s=1/4\pi$. 
The main information we get is that the correction we have is essentially $p_T$ independent especially for $v_2$
and this shows that changing $\sigma$ is not equivalent to change the $\eta/s$ that induce a quite $p_T$
dependent modification of $v_2(p_T)$. Therefore the two effect does not compensate each other when
looking at the $p_T$ dependence of the anisotropies.

\begin{figure}[h]
\centering
\includegraphics[width=1\linewidth]{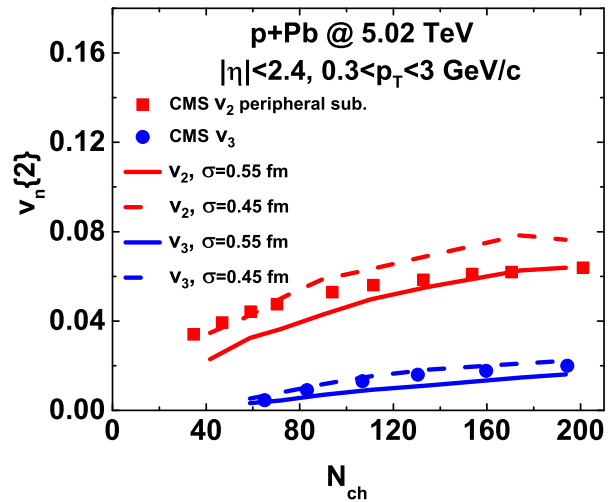}
\caption{(Color online) 
Integrated charged hadron $v_2$ and $v_3$ at $|\eta|<2.4$ and $0.3<p_T<3$ GeV$/c$ as a function of multiplicity in 5.02 TeV p+Pb collisions for two different choices of Gaussian width: $\sigma=0.55$ and 0.45 fm, compared to data from the CMS Collaboration~\cite{Chatrchyan:2013nka}.}
\label{fig:Nchvn}
\end{figure}

We also report in Fig. \ref{fig:Nchvn} that doing a study of $v_2$ and $v_3$ as a function of the multiplicity
 of charged hadrons at $|\eta|<2.4$ and $0.3<p_T<3$ GeV$/c$
 we find a good agreement also for the integrated $v_n$, but  in the most central collisions while we tend to underestimate $v_2$ in more peripheral collisions.
We find that both $v_2$ and $v_3$ increase with the increase of charged particle multiplicity. Because the eccentricity and size of the p+Pb systems change less than 6\% from 0-0.3\% to 30-40 \% centrality, the only reason for this is the shorter lifetime of low multiplicity events due to a lower initial energy density. 

However a small change of  $\sigma$ from 0.55 fm to 0.45 fm, allows to reproduce the $v_2$ in low multiplicity events, while we overestimate it in high multiplicity events. Changing $\sigma$ from 0.55 fm to 0.45 fm doesn't change 
significantly the integrated $v_3$, which can be seen by the relative small differences between solid and dashed blue lines in Fig.~\ref{fig:Nchvn}.
A change in $\sigma$ as a function of centrality does not have a solid physical motivation, we only wanted to convey the message that the quantitative agreement can be to some extent
tuned by the size of the spatial fluctuations. 
A real understanding of $pA$ asks in the future at least to extend the study to higher harmonics and their correlations as done
for $AA$ collisions.

\section{The impact of minijet at mid-$p_T$}

In the above section, we find that our results on spectra of charged hadrons at higher $p_T$ 
significantly underestimate experimental measurements, see Fig. \ref{fig:spectra}. 
However a main advantage of the transport approach with respect to the pure hydrodynamics is the possibility
to self-consistently include initial conditions that significantly deviate from the equilibrium one as the power law tail
associated to mini-jet production. In this section we extend our study including minijet at moderate mid transverse momentum,
however still limiting ourself to a region of $p_T <$ 5 GeV$/c$ because of limitations due to statistics but also because at higher
$p_T$ would be necessary to properly include also the radiative energy loss and the details that distinguish it
from an elastic collisional energy loss.
Because the non-equilibrium production of minijets in initial conditions can have an impact on both spectra and collective flows at higher $p_T$, we thus try to include them in initial conditions in this section. To do this, we use the spectra of gluons and quarks from CUJET Collaboration~\cite{Xu:2015bbz}  in 5.02 TeV p+p collisions, which is parametrized as 
\begin{equation}
\frac{dN_{g,q}}{d^2p_Tdy}=\frac{1}{120}(\frac{a}{1+p_T/b})^c (\rm{GeV}/c)^{-2},
\end{equation}
where $a=3.36479, b=1.53518$  GeV$/c$, $c=6.16767$ for gluons and $a=2.24291, b=1.31444$ GeV$/c$, $c=5.72108$ for three flavors light quarks (same for anti-quarks). The production of minijets in 5.02 TeV p+Pb collisions should be scaled with binary collisions, which can be obtained from the Monte Carlo Glauber model described above. For the distribution of minijets in transverse plane, we assume they are centered around the center of each binary collision pair, with the same Gaussian width as thermal partons. On the other hand, we assume minijets have no asymmetric distributions in spatial rapidity as they are produced by binary collisions. For the longitudinal momentum distribution, we assume that their rapidity is same as spatial rapidity because they are not in thermal equilibrium.

\begin{figure}[h]
\centering
\includegraphics[width=1\linewidth]{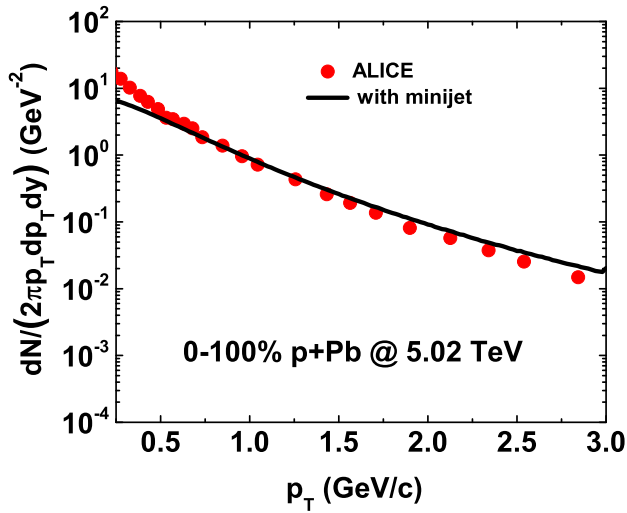}
\caption{(Color online) 
Charged hadron spectra at midrapidity in center of mass frame in minimum-bias 5.02 TeV p+Pb collisions with the inclusion of minijet at $p_T>$ 3 GeV$/c$ compared to data from the ALICE Collaboration~\cite{Abelev:2014dsa}.}
\label{fig:spectrajet}
\end{figure}

We have found that including minijets in initial conditions can lead to the enhancement of spectra at higher $p_T$, which can be seen by the solid black line in Fig.~\ref{fig:spectrajet}, where we include minijets at $p_T>3$ GeV$/c$. The reason for this is trivial, because exponential law decay of thermal partons decays faster compared to the power law decay of minijets.

\begin{figure}[h]
\centering
\includegraphics[width=1\linewidth]{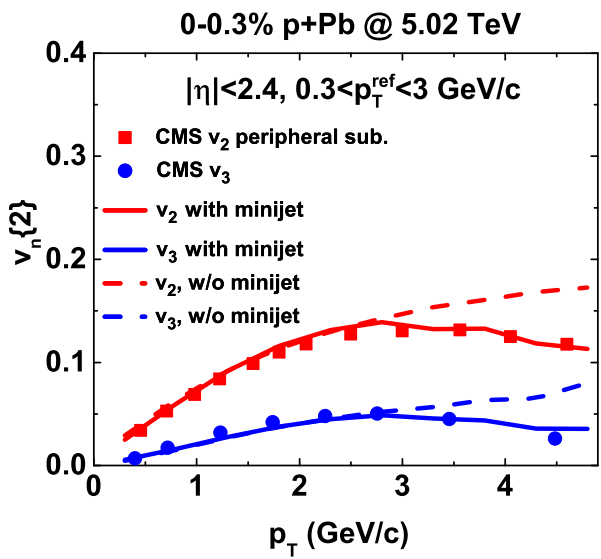}
\caption{(Color online) 
The transverse momentum $p_T$ dependence of $v_2$ and $v_3$ of charged hadrons at $|\eta|<2.4$ in 0-0.3\% 5.02 TeV p+Pb collisions with and without the inclusion of minijet at $p_T>$ 3 GeV$/c$, compared to data from the CMS Collaboration~\cite{Chatrchyan:2013nka}.}
\label{fig:vnjet}
\end{figure}

Moreover the mini-jets not only enhance the spectra of charged hadrons at higher $p_T$, but also can affect the transverse momentum dependence of collective flows. It is seen in Fig.~\ref{fig:vnjet} by the solid red and blue lines that the inclusion of minijets decreases $v_2$ and $v_3$ at $p_T \simeq$ 2.5 GeV$/c$, which leads to a better agreement with experimental data up to 5 GeV$/c$.
 
To our knowledge this is the first time the collective flows $v_{2,3}$ in $pA$ collisions are predicted correctly in such a
wide range of $p_T$. We think that the agreement with the experimental data shown in Fig. \ref{fig:vnjet} 
further strengthen the validity of the interpretation of the anisotropies observed as coming from an hydro-like expansion. We have to clarify that by hydro-like behavior we simply mean that we can achieve a reasonable description of collective anisotropies only assuming a matter expanding with a significantly highly rate of collisions similarly to a fluid. However, it remains an important open question if such an expanding phase is fully hydrodynamical or it is consistent with a transport evolution where non-hydrodynamical (particle-like) collective excitations are relevant. Very recently in Ref.~\cite{Kurkela:2019kip} a new quantitatively analysis, employing a conformal kinetic theory, has shown that $pA$ may lie at least in a region of transition between pure hydrodynamics and kinetic-particle evolution. It will certain be a study to be pursued to extend the study to our approach that contains also non conformal contribution to the expansion and could investigate also potential impact of the power-law tail in the distribution function.

\section{Flow correlations between $v_n$ and $\epsilon_n$ from large colliding systems to small ones}
The correlations between collective flows $v_n$ and initial asymmetry in coordinate space $\epsilon_n$ 
in $AA$ collisions have been studied in event-by-event hydrodynamics and transport approaches in recent years~\cite{Gardim:2011xv,Chaudhuri:2012mr,Niemi:2012aj,Plumari:2015cfa}. In general it has been shown that $v_2$ is strongly correlated with $\epsilon_2$ while higher flows are correlated weaker with $\epsilon_{n>2}$. In this section, we will 
study, to our knowledge for the first time in $pA$ collisions,  
compare such correlations to the one in $AA$ collisions.

To characterize the strength of the correlation between $v_n$ and $\epsilon_n$, we adopt the same correlation coefficient $C(n,m)$ defined in Ref.~\cite{Plumari:2015cfa}
\begin{equation}
C(n,m)=\frac{\sum_i(\epsilon_n^i-\langle \epsilon_n\rangle)(v_m^i-\langle v_m\rangle)}{\sqrt{\sum_i(\epsilon_n^i-\langle\epsilon_n\rangle)^2\sum_i(v_m^i-\langle v_m\rangle)^2}},
\end{equation}
where $\epsilon_n$ is defined as
\begin{equation}
\epsilon_n=\frac{\sqrt{\langle r_T^n\rm{cos} (n\phi)\rangle^2+\langle r_T^n \rm{sin}(n\phi)\rangle^2}}{\langle r_T^n \rangle}.
\end{equation}
$C(n,m)$ close to one corresponds to the stronger linear correlation between initial $\epsilon_n$ and final $v_m$.

\begin{figure}[h]
\centering
\includegraphics[width=1\linewidth]{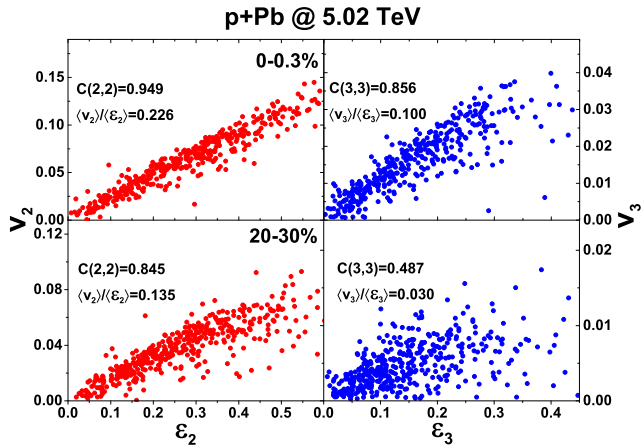}
\caption{(Color online) 
$\epsilon_n$ and $v_n$ in 0-0.3\% and 20-30\% 5.02 TeV p+Pb collisions for $n=2, 3$ with $\sigma=0.55$ fm.}
\label{fig:correlation}
\end{figure}

In Fig.~\ref{fig:correlation} we show the correlations between $v_2$, $v_3$ and $\epsilon_2$ and $\epsilon_3$ for 5.02 TeV p+Pb collisions at 0-0.3\% and 20-30\% centrality class with Gaussian width $\sigma=0.55$ fm. In the upper panel of Fig.~\ref{fig:correlation}, it is seen that the correlation between $\epsilon_2$ and $v_2$ is larger than the one between $\epsilon_3$ and $v_3$. From the central collision to peripheral collision, we find that the correlation decreases both for $n=2$ and $n=3$, though it decreases faster for the correlation between $\epsilon_3$ and $v_3$. This is a little different from the trend in Pb+Pb collisions, where there is almost no drop of the correlation for $n=2$, and only a small drop of it for $n=3$~\cite{Plumari:2015cfa}. In Fig.~\ref{fig:correlation}, we furthermore indicate the values of $\langle v_n\rangle/\langle\epsilon_n\rangle$ ratio, which decreases also for more peripheral collisions, same as the correlations.

We also studied the effect of varying the gaussian width $\sigma$ to see the effect on of  $C(n,n)$ and $\langle v_n\rangle/\langle\epsilon_n\rangle$ ratio in p+Pb collisions, and it is found that the their values are very similar for 
both the centrality class  0-0.3\% and 30-40\%. Because of this, we only compare our results in p+Pb collisions with $\sigma=0.55$ fm to what calculated in Ref.~\cite{Plumari:2015cfa} in Pb+Pb collisions.

\begin{figure}[h]
\centering
\includegraphics[width=1\linewidth]{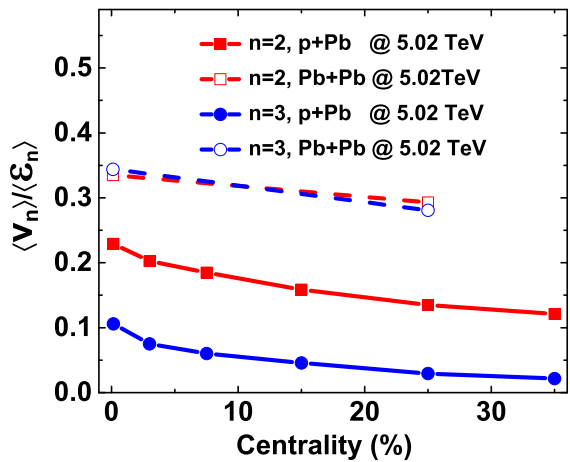}
\caption{(Color online) 
$\langle v_n\rangle/\langle\epsilon_n\rangle$ ratio as a function of centrality class in 5.02 TeV p+Pb collisions for $n=2$ and 3, compared to calculations in Ref.~\cite{Plumari:2015cfa} in 5.02 TeV Pb+Pb collisions.}
\label{fig:Nchve}
\end{figure}

In Fig.~\ref{fig:Nchve}, we show the $\langle v_n\rangle/\langle\epsilon_n\rangle$ ratio as a function of centrality class in 5.02 TeV p+Pb collisions for $n=2$ and 3 as well as those calculated  in Ref.~\cite{Plumari:2015cfa} in 5.02 TeV for Pb+Pb collisions. It is seen that both  $\langle v_2\rangle/\langle\epsilon_2\rangle$ and $\langle v_3\rangle/\langle\epsilon_3\rangle$ are quite smaller in p+Pb collisions compared to Pb+Pb collisions. Furthermore they all decrease at increasing of centrality, regardless of the size of colliding systems, but they decreases quite faster in small colliding systems.
From Fig. \ref{fig:Nchve} we can see that the smaller $v_{2,3}$ in p+Pb is due to a quite reduced efficiency in converting
the initial eccentricities $\epsilon_{n}$ even if the $\eta/s$ assumed is the same as the one for Pb+Pb. We notice that the value of $\langle v_2\rangle/\langle\epsilon_2\rangle$ and $\langle v_3\rangle/\langle\epsilon_3\rangle$, representing the efficiency of conversions of space anisotropy, are quite similar to those obtained in viscous hydrodynamics in Ref.~\cite{Nagle:2013lja} that however is a study at RHIC energy for p+Au, d+Au and $^3$He+Au.

\begin{figure}[h]
\centering
\includegraphics[width=1\linewidth]{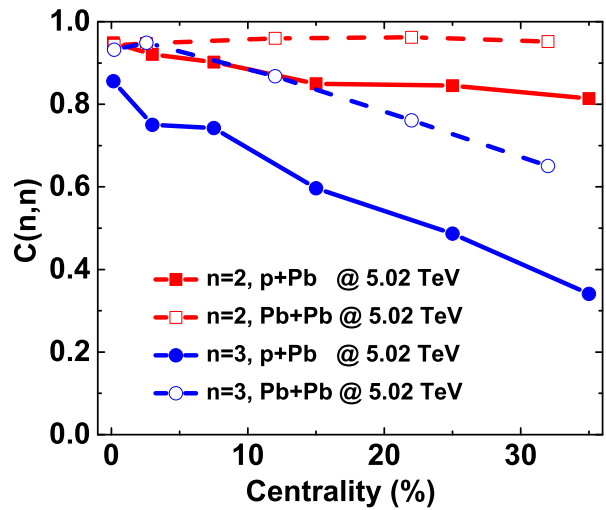}
\caption{(Color online) 
Correlation coefficient $C(n,n)$ as a function of centrality class in 5.02 TeV p+Pb collisions for $n=2$ and 3, compared to calculations in Ref.~\cite{Plumari:2015cfa} in 5.02 TeV Pb+Pb collisions.}
\label{fig:NchCnn}
\end{figure}

In order to better visualize the relation between the correlation coefficient $C(n,n)$ and centrality, we plot in Fig.~\ref{fig:NchCnn} the correlation coefficient as a function of centrality in 5.02 TeV p+Pb collisions for $n=2, 3$. It is seen by the solid red and blue lines that both $C(2,2)$ and $C(3,3)$ decreases with increase of centrality, while the correlation for triangular flow $C(3,3)$ is smaller than elliptic flow $C(2,2)$, and decreases faster. Compared to Pb+Pb collisions from Ref.~\cite{Plumari:2015cfa}, we find that the correlation coefficients for $n=2$ and $n=3$ are smaller for small colliding systems and decreases faster with centrality class.

We propose to access such correlations in $pA$ systems because their measurements could give an important contribution to
validate the interpretation of the anisotropies $v_n$ in $pA$ collisions as hydro-like collective expansion.

\begin{figure}[h]
\centering
\includegraphics[width=1\linewidth]{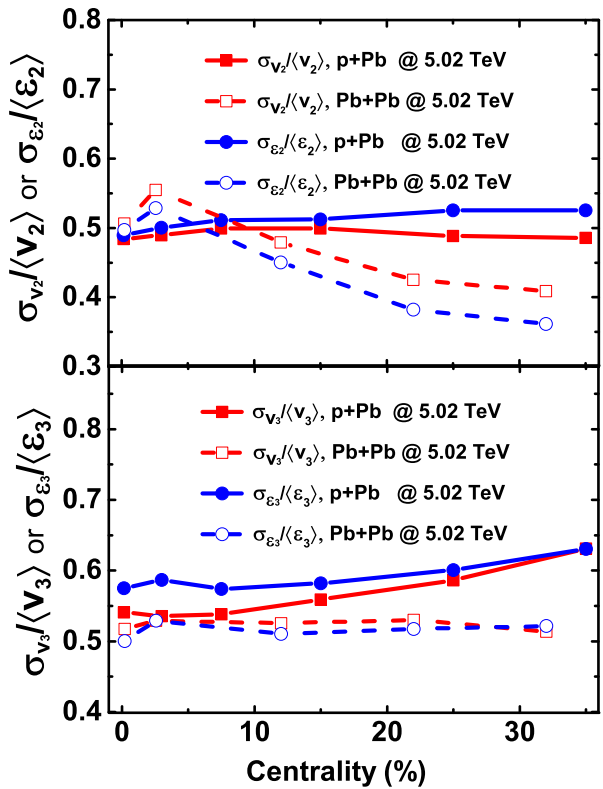}
\caption{(Color online) 
 $\sigma_{v_n}/\langle v_n\rangle$ and $\sigma_{\epsilon_n}/\langle\epsilon_n\rangle$ as a function of centrality class in 5.02 TeV p+Pb collisions for $n=2$ and 3, compared to calculations in Ref.~\cite{Plumari:2015cfa} in 5.02 TeV Pb+Pb collisions.}
\label{fig:Nchev}
\end{figure}

The correlation between collective flows and the initial geometry is strong, and the initial geometry of p+Pb collisions is dominated by fluctuations, the distribution of $v_n$ should be quite relevant as an indicator of initial conditions. 
Therefore we conclude our study of $v_{2,3}$ in $pA$ collisions discussing the normalized variance
 $\sigma_{v_n}/\langle v_n\rangle$. In Fig.~\ref{fig:Nchev} 
we plot the centrality dependence of  both $\sigma_{v_n}/\langle v_n\rangle$ and $\sigma_{\epsilon_n}/\langle \epsilon_n\rangle$ in 5.02 TeV p+Pb collisions, as well as the same one in 5.02 TeV Pb+Pb collisions from Ref.~\cite{Plumari:2015cfa}. Shown in solid and dashed blue lines in the upper panel of Fig.~\ref{fig:Nchev} for $n=2$, we find that $\sigma_{\epsilon_2}/\langle \epsilon_2\rangle$ increases slightly with centrality class in p+Pb collisions. 
In Pb+Pb the decrease with centrality class can be attributed to 
the additional contribution of $\epsilon_2$ from the global average geometry in latter case. 
According to our modeling on intial conditions in $pA$ $\epsilon_n$ is dominated by fluctuations at all centralities
entailing a $\sigma_{v_2}/\langle v_2\rangle$ that even slightly increase with centrality 
as shown by the solid red line, compared to the decrease of it with centrality in large colliding systems 
as shown by the dashed red line.
Of course observing such a different trend would give further support to the current modeling of the initial conditions.

In the lower panel of Fig.~\ref{fig:Nchev}, we show $\sigma_{v_3}/\langle v_3\rangle$ and $\sigma_{\epsilon_3}/\langle \epsilon_3\rangle$ as a function of centrality class in p+Pb and Pb+Pb collisions. It is seen by the solid and dashed red lines that $\sigma_{v_3}/\langle v_3\rangle$ increases with centrality class in p+Pb collisions while stays almost constant in Pb+Pb collisions, which indicates that triangular flow is dominated by fluctuations in both small and large colliding systems, but
anyway there is significant tendency to stay larger in $pA$ systems. We also find that  $\sigma_{\epsilon_3}/\langle \epsilon_3\rangle$ is smaller in large colliding systems compared to small ones. Shown by the solid and dashed red lines, $\sigma_{v_3}/\langle v_3\rangle$ also increases slowly with centrality class in p+Pb collisions, while it stays the same in Pb+Pb collisions and below the value in p+Pb collisions, which agrees with the trend of $\sigma_{\epsilon_3}/\langle \epsilon_3\rangle$.
We think that such a measurements can significantly contribute to validate or falsify our modeling of
the initial conditions and of the dynamics of $pA$ collisions.

\section{Conclusions and Discussions}
We use an event-by-event transport approach whose initial conditions are generated by wounded quark model as well as negative binomial distribution overlaid on partons production for each participant quark, to study several hadron observables in p+Pb collisions. We find that we can reproduce quite well $dN/d\eta$, the distribution of charged particle multiplicity and the spectra of charged hadrons up to 1.5 GeV$/c$ at midrapidity. 
We have shown that the anisotropies $v_2$ and $v_3$ are building-up in time in a way that roughly scales with
the mean square radius of the system. So the formation time in $pA$ system will be in general about a factor 4 faster
with respect to most central $AA$ collisions.
A main result is that the transverse momentum dependence of $v_2$ and $v_3$ predicted in our approach
agree with experimental measurements with $\eta/s=1/4\pi$ up to about 2.5 GeV$/c$, as well as the integrated $v_2$ and $v_3$. 
This is in general confirm several results already obtained in viscous hydrodynamics \cite{Bozek:2011if,Bozek:2013uha,Kozlov:2014fqa,Weller:2017tsr,Shen:2016zpp}, we also find that however ultra-central p+Pb collisions are more sensitive to
the value of the $\eta/s$ assumed.
A specific advantage of our approach is the possibility to include also power law tail associated to mini-jets production,
we have found that this allows to extend the 
 agreement with experimental data on  collective flows at higher $p_T$ at least up to
 about 5 GeV$/c$ and this is also accompanied by a much better prediction of the transverse momentum spectrum.
We have also pointed out that at variance with respect to Pb+Pb in small systems the $\epsilon_n$ and $v_n$
are strongly dependent on the width of the initial spatial gaussian fluctuations. There is some interplay between
the value of the $\eta/s$ and the width of the initial fluctuations, however especially for $v_2$ reducing the width
induces an increase of the elliptic flow that is $p_T$ independent.

Finally, we furthermore study the correlations between $v_n$ and $\epsilon_n$ that has been intensively studied for
$AA$ collisions, but to our knowledge they are for the first time discussed for $pA$ collisions. 
In general, we find that the correlation coefficient is still strong for $n=2$ in 0-0.3\% collisions similarly
to $AA$ collisions, but it becomes weaker for either the higher harmonics $v_3$ or larger centrality class, and the correlation coefficient decreases quite faster with centrality in p+Pb collisions compared to Pb+Pb collisions. So we can say that except
ultra-central collisions and only for $v_2$ we expect significantly less correlations in p+Pb collisions.
Besides that, we further predict that except for ultra-central collisions the variance $\sigma_{v_2}/\langle v_2\rangle$ should be nearly centrality independent and quite larger than the one in Pb+Pb.
We plan in a further study to investigate if an initial glasma phase can affect such behaviors. In the meantime
experimental measurements can shed new light on the goodness of the present modeling of $pA$ as liquid drops
with local density fluctuations expanding nearly hydrodynamically.

\section*{ACKNOWLEDGEMENTS}
We gratefully acknowledge useful discussions with Jamie L. Nagle. The work of  Y.S. is supported by a INFN post-doc fellowship within the national SIM project.

\bibliography{ref.bib}

\end{document}